# Strain-Induced Band Engineering in Monolayer Stanene on Sb(111)

Jian Gou[1,3], Longjuan Kong[1,3], Hui Li[2], Qing Zhong[1,3], Wenbin Li[1,3], Peng Cheng[1,3], , Lan Chen[1,3*] and Kehui Wu[1,3,4*]

[1]*Institute of Physics, Chinese Academy of Sciences, Beijing 100190, China*
[2]*Beijing Advanced Innovation Center for Soft Matter Science and Engineering, Beijing University of Chemical Technology, Beijing 100029, China*
[3]*School of physics, University of Chinese Academy of Sciences, Beijing 100049, China*
[4]*Collaborative Innovation Center of Quantum Matter, Beijing 100871, China*
*Email: lchen@iphy.ac.cn (L. C.), khwu@iphy.ac.cn (K. W),

**Abstract:**

Two-dimensional (2D) allotrope of tin with low buckled honeycomb structure, named as stanene, is proposed to be an ideal 2D topological insulator with a nontrivial gap larger than 0.1 eV. Theoretical works also pointed out the topological property of stanene occurs by strain tuning. In this letter, we report the successful realization of high quality, monolayer stanene film as well as monolayer stanene nanoribbons on Sb(111) surface by molecular beam epitaxy, providing an ideal platform to the study of stanene. More importantly, we observed a continuous evolution of the electronic bands of stanene across a nanoribbon, which are related to the strain field gradient in stanene. Our work experimentally confirmed that strain is an effective method for band engineering in stanene, which is important for fundamental research and application of stanene.

Two dimensional (2D) topological insulators (TIs) are featured by an energy gap in the interior and nontrivial, gapless states along the edge that can be described by the Dirac equation [1, 2]. Due to the time-reversal-symmetry (TRS), the spin and momentum degrees of freedom of the edge states are locked. The backscattering of carriers, which would require a spin-flip process, is not allowed in 2D TIs if a time-reversal invariant perturbation, such as nonmagnetic disorder, is present. These intriguing properties are the basis of the topological protection, and the expected helical spin-polarized transport makes 2D TIs potentially very promising materials that could serve as a platform for realizing quantum spin Hall effect (QSHE) [1, 3], quantum computing [4] and spintronic device applications [5].

The first 2D TI proposed by theorists is graphene [1], but the extremely small energy gap (about $10^{-3}$ meV) opened by spin-orbit coupling (SOC) makes it unlikely to observe any real effect in the experimental accessible temperature range [6]. In 2007, QSHE was first observed in an HgTe quantum well [7]. However, fabrication of HgTe quantum well is very difficult and successful experiment based on HgTe is very limited. Therefore, searching for new 2D TI is desirable. Compared with graphene, honeycomb structures formed by other group-IV elements, such as silicene and germanene have larger spin-orbit coupling strength, which are predicted to lead to larger energy gaps of 1.55 meV and 23.9 meV, making the $Z_2$ TIs measurable under experimentally achievable temperatures [8]. Furthermore, Xu and Zhang *et al.* predicted that stanene with low-buckled honeycomb structure can have an intrinsic SOC-induced gap as large as 100 meV, suggesting possible room temperature QSHE [14]. Moreover, by chemical functionalization of stanene, a parity exchange between occupied and unoccupied bands occurs at the Γ point and creates a nontrivial bulk gap of up to 0.3 eV. Remarkably, such a large bulk gap is strongly associated with the in-plane strain induced by surface functionalization [14].

Considering the fact that stanene is presently the most promising, simplest 2D TI material, it is very important to systematically investigate its structure and electronic properties, especially the possible effect of engineering its electron band structure under controllable strain. Recently, the first experimental realization of stanene was reported on $Bi_2Te_3$, but the growth of tin on $Bi_2Te_3$ adopts a Vollmer-Weber growth mode (island growth) [13]. As a result, it is difficult to obtain a uniform monolayer stanene on the substrate, and thus the experimental investigation on the strain effect on the electronic structures of stanene is still very limited so far.

In this Letter, reported the successful growth of uniform monolayer stanene on Sb(111) surface by molecular beam epitaxy (MBE). Combined with scanning tunneling microscopy/spectroscopy (STM/STS) and first-principle calculations, we have revealed that monolayer stanene on Sb(111) has a compressed honeycomb lattice. Moreover, we have demonstrated a strain-induced electron band engineering effect in stanene nanoribbons. The electronic bands corresponding to $P^+_{x,y}$ orbitals of Sn move to Fermi level gradually along the transverse direction across nanoribbon, which are proven to be a result of strain relaxation across the stanene ribbon. These results pave the way to future applications of stanene in electronic and spintronic devices.

The strategy to choose Sb(111) as substrate is that antimony is a layered semimetal with ABC staking sequence along the [111] crystallographic direction. Every biatomic Sb layer can be considered as buckled honeycomb lattice similar to stanene, with weak van der Waals interaction between layers (as a result, Sb(111) surface can be easily obtained by cleavage [15, 16]). More importantly, previous works on Sb/Sn binary alloy revealed a structure of intercalating pure Sb and Sn layers along the hexagonal *c*-axis [17, 18], suggesting that a layer-by-layer growth of Sn on Sb is possible.

The experiments were carried out in a home-built ultrahigh vacuum STM/MBE system. The single crystal Sb(111) surface was cleaned by standard $Ar^+$ ion sputtering and annealing process. Sn with the purity of 99.999% was evaporated from a resistance-heating crucible. During the deposition, Sb(111) substrate was kept at about 400 K in the MBE chamber with background pressure better than $2\times10^{-10}$ Torr. After Sn deposition, the sample was transferred *in-situ* to the STM chamber. All STM and STS measurements were performed at liquid helium temperature by a chemically etched tungsten tip. The differential conductance (dI/dV) spectra were measured by the in-plane ac component in the tunneling current with a lock-in amplifier by superimposing an ac voltage (10 mV, 669 Hz) on the given dc bias of the substrate-tip gap. The STM topographic images were processed by WSxM [19].

First-principles density functional theory calculations were carried out with the Vienna *ab initio* Simulation Package (VASP) [20]. In the present calculations, the interaction between valence and core electrons was described by the projector-augmented wave (PAW) and the exchange-correlation interaction was treated using the generalized gradient approximation (GGA) in the formulation of Perdew–Burke–Ernzerhof (PBE) function [21, 22]. The plane-wave cutoff energy was set to 500 eV and the vacuum space was set to be ~ 19 Å. A biatomic tin film was placed on a three-layer 1×1 Sb (111) slab with fixed bottom layer to mimic the semi-infinite solid. The uppermost surface layers and stanene were fully relaxed until the residual force on each atom was less than 0.01 eV/ Å.

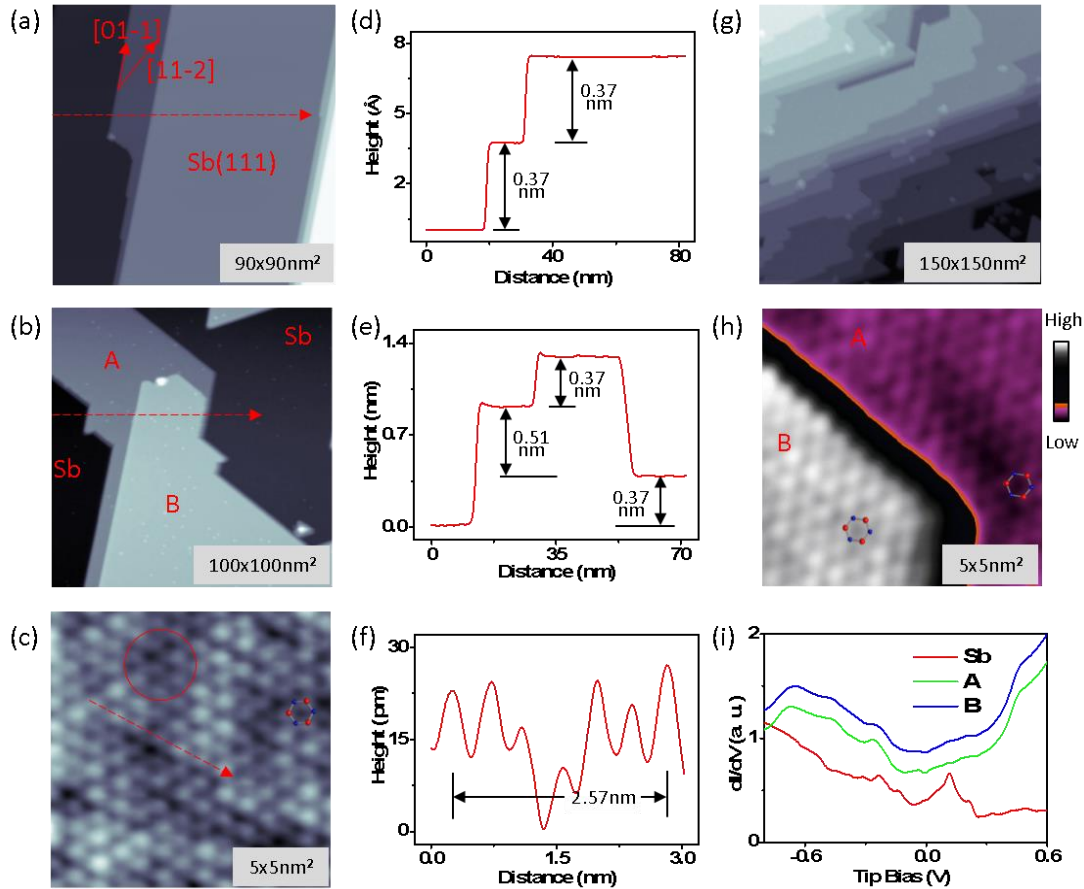

FIG. 1. Single Sn layer grown on Sb(111). (a) STM topographic image (-2.2 V, 90 pA) of clean Sb(111) surface with straight step edges. Crystallographic axes are labeled with red arrows. (b) STM topographic image (-2.0 V, 90 pA) of 0.5 ML tin grown on Sb(111). (c) High-resolution STM image of stanene (0.40 V, 100 pA). Depression area marked by red circle illustrates a clear honeycomb-like atomic structure. (d),(e),(f) Height line profiles along the dashed arrows in (a),(b),(c), respectively. The step heights are indicated. (g) Large scale STM topographic image of stanene with a coverage of 0.95 ML (-2.4 V, 90 pA). (h) High-resolution STM image containing terraces of island 'B' and 'A' (0.4 V, 0.3 nA). Two ball-and-stick models in which red and blue balls represent upper bucked and lower bucked Sn, are superimposed on the protrusions to indicate the atomic structure. (i) dI/dV spectra taken on a clean Sb(111) terrace (red), island A (green) and B (blue), respectively. The curves are offset vertically for clarity.

The typical STM image of Sb(111) surface is shown in Fig. 1(a), in which the crystallographic directions of Sb(111) can be determined by the straight steps. The line profile across the steps (Fig. 1(d)) reveals a step height of about 0.37 nm, consistent with the thickness of single Sb layer. Fig. 1(b) is STM image of the Sb(111) surface after deposition of about 0.5 ML Sn. Sn islands with triangular and strip shapes are formed around the step edges of the Sb(111) substrate, and the edges of Sn islands are along the same high-symmetry crystallographic orientations of Sb(111). Another feature in Fig. 1(b) is that two different layers of Sn seem to be formed on the Sb(111) surface: the first layer (label A) and second layer (label B). But from the line profile across steps of Sb(111) and Sn islands (Fig.1(e)), two different heights appear. The first Sn layer (A) has a height of 0.51±0.01 nm relative to Sb substrate, while the height difference between B and A is 0.37±0.01 nm, which is same as the height of single Sb layer. Therefore, we believe island B is not a second Sn layer, but the same first Sn layer on a Sb terrace higher than that below island A. The similar phenomena can also be found in FeO/Pt(111) growth [23]. Moreover, the STS measurements on terraces of A and B reveal the same curve of local density of states (LDOS) (Fig. 1(i)), proving they are the same Sn layer.

The atomic structure of the first Sn layer is shown in the high-resolution STM image in Fig. 1(c), in which we found the brightness of protrusions is inhomogeneous. In the depression area marked by the red circle a buckled honeycomb structure can be clearly revealed, suggesting that the Sn layer is monolayer stanene structure. The line profile (Fig. 1(f)) along the red dashed arrow shown in Fig. 1(c) indicates a similar lattice constant (0.43±0.01 nm) with substrate which is slightly smaller than either Sn(111) surface [24] or freestanding stanene (0.468 nm) [14]. The smaller lattice constant suggest the existence of compressive strain in stanene on Sb(111), which may cause a slight fluctuation in the buckling degree of Sn atoms, and it explains the inhomogeneity in the image contrast in Fig. 1(c). Additionally, the high-resolution STM image containing both A and B terraces shown in Fig. 1(h) reveals the same honeycomb structures and identical crystallographic orientations, giving another evidence that both of them are monolayer stanene. Increasing tin coverage, a completed monolayer stanene can be formed on the entire substrate surface, as shown in Fig. 1(g). Further deposition of Sn atoms on Sb(111) with coverage exceeding one monolayer will form 2×2 reconstruction on the top of Sn layer, which is similar to the α-Sn(111) grown on InSb(111) substrate [24, 25]. Therefore, the growth of tin on Sb(111) should be layer-by-layer mode.

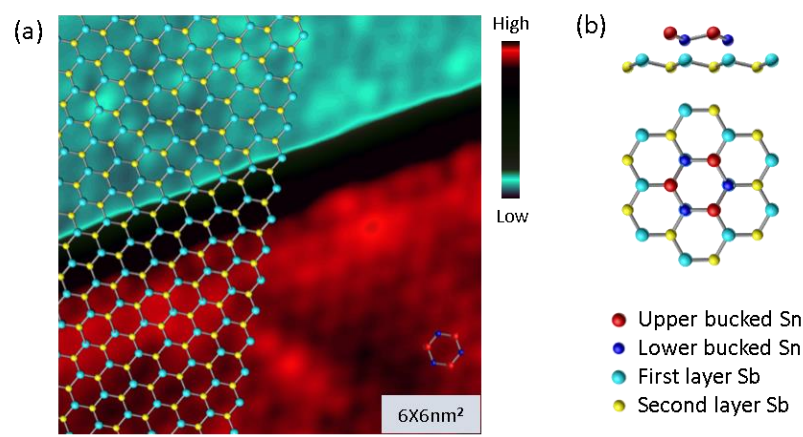

Fig. 2. The atomic structure and staking mode of stanene on Sb(111). (a) Atomic-resolution STM image of stanene (lower part) on Sb(111) substrate (upper part) (0.4 V,0.3 nA). Honeycomb model, in which the yellow and cyan balls represent the structure of the topmost Sb layer, is superimposed and extend from Sb to stanene to illustrate the staking relationship between them. (b) Top view (upper panel) and side view (lower panel) of the stacking model for tin film on Sb(111).

Next, we have determined the stacking configuration of stanene on Sb(111) substrate through high-resolution STM image. Fig. 2(a) gives an atomic-resolution image where the atoms on stanene and the Sb(111) substrate are resolved simultaneously, thus the relative position of the stanene lattice and the Sb(111) lattice can be determined. According to crystallographic orientation of Sb(111) (Fig. 1(a)), a ball-and-stick model of Sb(111) surface is superimposed and extend from Sb to stanene (bottom right) to illustrate the stacking relationship. Unambiguously, the stacking relationship between stanene and Sb(111) surface is determined to be a *AA'* stacking sequence, in which the lower bucked Sn atoms locate at the topmost site of Sb and upper bucked Sn atoms locate at the sites above the second layer Sb atoms. The *AA'* stacking sequence in stanene on Sb(111) is different to the germanene on Sb(111), which adopts *AB* stacking configuration [26]. The structural model of stanene on Sb(111) is shown in Fig. 2(b).

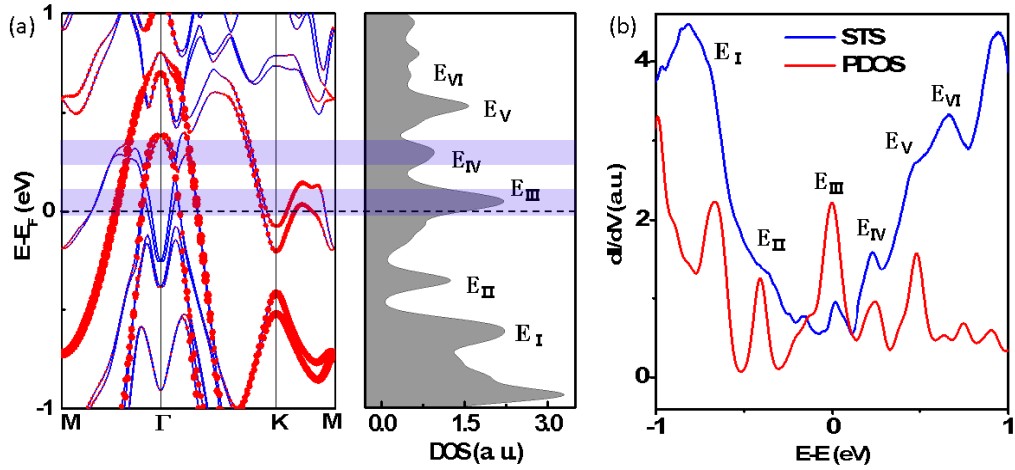

Fig. 3. The electronic structure of stanene on Sb(111). (a) Left panel: calculated band structure with SOC. Blue curves stand for the total band structure including three layers of biatomic Sb(111) substrate and a single layer stanene on it. Red dots indicate the projected bands on stanene and the size of dos represents the weights factor. Right panel: the DOS of stanene. (b) Comparison of experimental dI/dV spectra (blue curve) and calculated DOS (red curve) of stanene on Sb(111). Energy positions of all the calculated DOS maximum agree with peaks in the experimental data very well. Several featured DOS maxima are labelled with $E_i$ (i=I, II, III, IV, V, VI).

The stacking configuration of stanene adlayer on Sb(111) are reconfirmed by first-principle calculations on infinite stanene monolayer adsorbed on Sb(111). The calculation results indicate that *AA'* stacking is an energy favorable structure for stanene on Sb(111). Based on such a compressed honeycomb stanene layer with *AA'* stacking on Sb(111), the calculated electronic structure is plotted in Fig. 3(a). The total electronic bands of stanene/Sb(111) with SOC and the projected bands on stanene adlayer are shown in the left panel. Remarkably, a gap of ~200 meV is opened at the K point, which is about twice of that in freestanding stanene. This demonstrates the significant effect of the compressive strain in stanene on Sb(111) (-8% lattice constant change relative to freestanding form). Meanwhile, the electronic band consisting of $P^+_{x,y}$ orbitals of tin atoms near the Γ point cross over the Fermi level and make stanene metallic, which is similar to that in stanene/$Bi_2Te_3$ [13].

The corresponding DOS of stanene is shown in the right part of Fig. 3(a), and the DOS maxima are labelled with $E_i$ (i=I, II, III, IV, V, VI) for identification. Compared with the band structure, the $E_{III}$ and $E_{IV}$ peaks can mainly be attributed to the van Hove singularities of $P^+_{x,y}$ band at the Γ point, and the flat bands at K-M, respectively. In other words, these two peaks are pure bands of stanene. On the contrary, other peaks are mainly contributed by the hybridized Sn orbitals with underling Sb atoms, except for $E_V$ which is contributed by $P^+_{x,y}$ band partly. The comparison of the calculated DOS and experimental STS curve is shown in Fig. 3(b). We found the peaks in STS curve correspond to the DOS maxima very well, except for a slight energy shift of less than 50 meV. The perfect matching between calculated DOS and STS measurements strongly supports our structural model for monolayer stanene on Sb(111). On the other hand, the slight energy shift may result from n-type doping which comes from surface defects. Similar doping effect was also found in other 2D materials with honeycomb structures [27, 28].

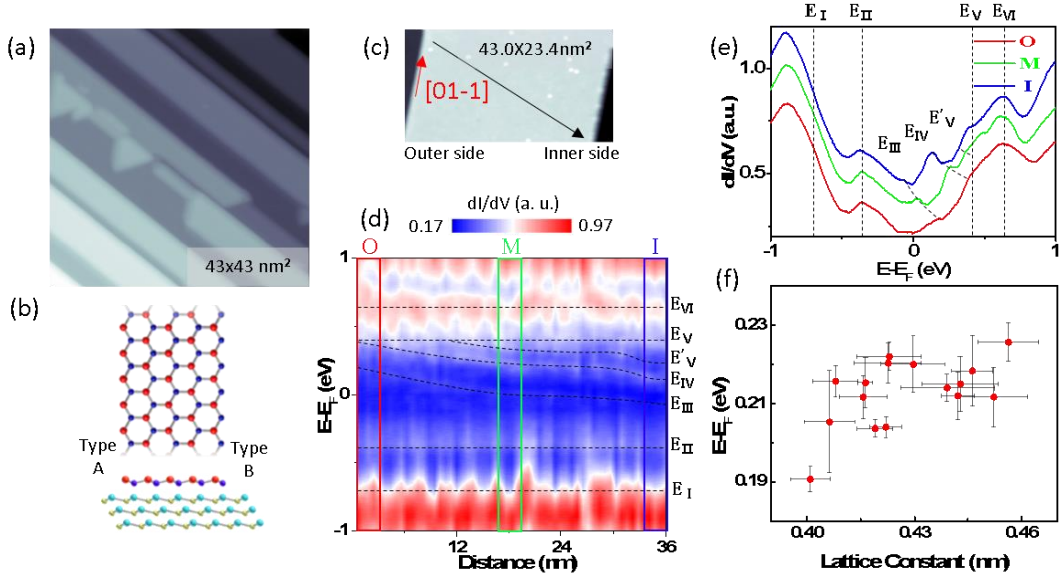

Fig. 4. The strain induced bands engineering. (a) STM topographic image of stanene nanoribbon on narrow substrate terraces (-1.5 V, 90 pA). (b) Schematic atomic structure of stanene nanoribbon on Sb(111). Upper panel: top view (only stanene was shown). Lower panel: side view. (c) STM image of a stanene nanoribbon (-1.0 V, 87 pA). The crystallographic orientation of Sb(111) are labelled with red arrow. (d) The color map of dI/dV spectra obtained along the line indicated by the black arrow in (c). (e) Average of the dI/dV spectra in rectangles labeled by the same colored 'O', 'M', 'I' in (d), respectively. The curves are offset vertically for clarity. Dashed lines marked by $E_i$ (i=I, II, III, IV, V, VI) in (d) and (e) are added along the featured STS peaks to guide the eye. (f) The statistics of the energy position dependence of $E_{III}$ on lattice constant of stanene.

Since the growth of stanene islands starts from the step edges of Sb(111) surface, we can obtain stanene nanoribbons on Sb(111) surface with narrow terrace arrays, as shown in Fig. 4(a). We performed STS measurements on a stanene nanoribbon shown in Fig.4(c) along the transverse direction across the nanoribbon. As illustrated in Fig. 4(d), a set of dI/dV spectra were depicted as a function of distance along the highest crystallographic orientation indicated by the black arrow in Fig. 4(c). From outer side edge to inner side edge, one can explicitly see an $E'_V$ peak split from $E_V$, and the positions of $E_{III}$ and $E_{IV}$ peaks move to Fermi level independently. On the contrary, other peaks are kept unchanged.

As we mentioned before, the $E_{III}$ and $E_{IV}$ peaks are pure bands of stanene. These bands are derived from the $P^+_{x,y}$ orbitals of tin, which are sensitive to in-plane strain according to previous theoretical works [14]. On the other hand, based on the stacking configuration of stanene on Sb(111), and the STM observation that the edges of stanene nanoribbons are along the crystallographic orientation of Sb(111) (Fig. 4(c)), we can conclude that the edges of nanoribbons are zigzag-like. Due to the buckled structure of stanene, there are two kinds of zigzag edges: type A (outermost atoms are upper buckled) and type B (outermost atoms are lower buckled), shown in Fig. 4(b). The stanene nanoribbon interacts with the Sb(111) substrate mainly through its edge atoms as these atoms have unsaturated dangling bonds. The type A edge should have a weaker interaction with Sb(111) substrate compared with type B edge, because the upper buckled tin atoms have further distance from substrate. In consequence, the compressive strain at the outer edge is relatively small and will increase gradually to inner side along the transverse direction.

The evolution of $E_{III}$ and $E_{IV}$ peaks from outer to inner edge of a stanene nanoribbon proves that the band structure of stanene can be tuned by strain. This is remarkable as further band engineering, such as band crossing or nontrivial gap opening may be achieved [14]. Considering the $E_V$ peak partly consisting of $P^+_{x,y}$ orbitals at $\Gamma$ point (Fig. 3(a)), which are sensitive to in-plane strain. The evolution of strain also induces electronic bands corresponding to $P^+_{x,y}$ orbitals in $E_V$ peak moving to Fermi level, which results in a $E'_V$ peak splitting from $E_V$ peak and movement. The previous theoretical work suggest the $P^+_{x,y}$ orbitals splitting at $\Gamma$ point due to the SOC effect for Sn-Sn bonds, which attribute to the $E_{IV}$ and $E'_V$ peaks in our experiments. The interval between the splitting $P^+_{x,y}$ orbitals can be very slightly enlarged under hydrostatic strain, which accorded with the unchanged interval between $E_{IV}$ and $E'_V$ (Fig. 4(d)) from outer side to inner side very well, proving existence of the strain induced band engineering.

Meanwhile, the strain-induced band tuning effect can also be revealed in larger stanene islands. For example, we performed STS measurements at various sites on the terrace of stanene islands with fluctuation of buckling degree of Sn atoms in Fig. 1(c). The energy position of the $E_{III}$ peak in dI/dV spectra shows an obvious monotonous dependence on the measured local lattice constant, as shown in Fig. 4(f). As the local lattice constant is directly related with the in-plane stain, this once again verifies the strain-induced band engineering effect in stanene.

Although free-standing stanene is a large gap 2D TI, stanene on different substrates, such as $Bi_2Te_3$ [13] so far all suffer from the interaction between stanene and the substrate, and becomes topological trivial systems [29]. It is therefore crucial to restore its band crossing and topological nature, for example by strain engineering, in order to enable novel applications. Stanene on Sb(111) provides an ideal stanene system, as monolayer stanene can be uniformly fabricated in large scale. The effective band engineering by varying in-plane strains implies that this is a promising way. As antimony is a layered material that can be exfoliated vastly, growing stanene on ultra-thin antimony film and combined with halogenations adsorption may further enhance the strain engineering effect, and even to realize a nontrivial gap opening in stanene. Finally, single layer antimony, namely antimonene, is semiconducting, whose structure is easy to be tuned [31, 32]. Combing stanene/antimonen offers a clean, interesting Van der Waals heterostructure system, whose properties are worth for further investigations [33].

**Acknowledgement**

We thank Dr. Yong Xu of Tsinghua University for constructive discussion. This work was supported by the MOST of China (grants nos. 2016YFA0300904, 2016YFA0202301, 2013CBA01601, 2013CB921702), the NSF of China (grants nos. 11674366, 11674368, 11334011, 11304368), and the Strategic Priority Research Program of the Chinese Academy of Sciences (grant no. XDB07020100).

**References:**

[1]   C. L. Kane and E. J. Mele, Phys Rev Lett **95**, 226801 (2005).
[2]   B. A. Bernevig, T. L. Hughes, and S.-C. Zhang, Science **314**, 1757 (2006).
[3]   X.-L. Qi and S.-C. Zhang, Physics Today **63**, 33 (2010).
[4]   C. Nayak, S. H. Simon, A. Stern, M. Freedman, and S. Das Sarma, Reviews of Modern Physics **80**, 1083 (2008).
[5]   D. Pesin and A. H. MacDonald, Nat. Mater. **11**, 409 (2012).
[6]   Y. Yao, F. Ye, X.-L. Qi, S.-C. Zhang, and Z. Fang, Phys. Rev. B **75** (2007).
[7]   M. König, S. Wiedmann, C. Brüne, A. Roth, H. Buhmann, L. W. Molenkamp, X.-L. Qi, and S.-C. Zhang, Science **318**, 766 (2007).
[8]   C.-C. Liu, W. Feng, and Y. Yao, Phys. Rev. Lett. **107** (2011).
[9]   P. Vogt, P. De Padova, C. Quaresima, J. Avila, E. Frantzeskakis, M. C. Asensio, A. Resta, B. Ealet, and G. Le Lay, Phys. Rev. Lett. **108** (2012).
[10]  L. Chen, C.-C. Liu, B. Feng, X. He, P. Cheng, Z. Ding, S. Meng, Y. Yao, and K. Wu, Phys. Rev. Lett. **109** (2012).
[11]  M. Derivaz, D. Dentel, R. Stephan, M. C. Hanf, A. Mehdaoui, P. Sonnet, and C. Pirri, Nano Lett. **15**, 2510 (2015).
[12]  C.-C. Liu, H. Jiang, and Y. Yao, Phys. Rev. B **84** (2011).
[13]  F. F. Zhu, W. J. Chen, Y. Xu, C. L. Gao, D. D. Guan, C. H. Liu, D. Qian, S. C. Zhang, and J. F. Jia, Nat. Mater.   (2015).
[14]  Y. Xu, B. Yan, H. J. Zhang, J. Wang, G. Xu, P. Tang, W. Duan, and S. C. Zhang, Phys Rev Lett **111**, 136804 (2013).
[15]  B. Stegemann, C. Ritter, B. Kaiser, and K. Rademann, Phys. Rev. B **69** (2004).
[16]  J. Seo, P. Roushan, H. Beidenkopf, Y. S. Hor, R. J. Cava, and A. Yazdani, Nature **466**, 343 (2010).
[17]  V. Vassiliev, M. Lelaurain, and J. Hertz, Journal of Alloys and Compounds **247**, 223 (1997).
[18]  L. Norén, R. L. Withers, S. Schmid, F. J. Brink, and V. Ting, Journal of Solid State Chemistry **179**, 404 (2006).
[19]  I. Horcas, R. Fernandez, J. M. Gomez-Rodriguez, J. Colchero, J. Gomez-Herrero, and A. M. Baro, The Review of scientific instruments **78**, 013705 (2007).
[20]  G. Kresse and J. Furthmüller, Physical Review B Condensed Matter **54**, 11169 (1996).
[21]  J. P. Perdew, K. Burke, and M. Ernzerhof, Phys. Rev. Lett. **77**, 3865 (1996).

[22]  P. E. Blöchl, Phys. Rev. B **50**, 17953 (1994).
[23]  W. Weiss and M. Ritter, Phys. Rev. B **59**, 5201 (1999).
[24]  T. Osaka, H. Omi, K. Yamamoto, and A. Ohtake, Phys. Rev. B **50**, 7567 (1994).
[25]  T. Eguchi, J. Nakamura, and T. Osaka, J. Phys. Soc. Jpn. **67**, 381 (1998).
[26]  J. Gou, Q. Zhong, S. Sheng, W. Li, P. Cheng, H. Li, L. Chen, and K. Wu, 2D Materials **3**, 045005 (2016).
[27]  L. Zhang, P. Bampoulis, A. N. Rudenko, Q. Yao, A. van Houselt, B. Poelsema, M. I. Katsnelson, and H. J. Zandvliet, Phys. Rev. Lett. **116**, 256804 (2016).
[28]  S. Jung, G. M. Rutter, N. N. Klimov, D. B. Newell, I. Calizo, A. R. Hight-Walker, N. B. Zhitenev, and J. A. Stroscio, Nature Physics **7**, 245 (2011).
[29]  At least, we didn't find any obviously conducted edge states at the outer zigzag side.
[30]  E. S. Walker *et al.*, Nano Lett. **16**, 6931 (2016).
[31]  X. Wu *et al.*, Adv. Mater. **29** (2017).
[32]  M. Zhao, X. Zhang, and L. Li, Sci. Rep. **5**, 16108 (2015).
[33]  A. K. Geim and I. V. Grigorieva, Nature **499**, 419 (2013).